# A Hybrid Deep Learning-Based State Forecasting Method for Smart Power Grids

Shahrzad Hadayeghparast
*School of Engineering*
*University of Guelph*
Guelph, Canada
shadayeg@uoguelph.ca

Amir Namavar Jahromi
*School of Engineering*
*University of Guelph*
Guelph, Canada
anamavar@uoguelph.ca

Hadis Karimipour
*School of Engineering*
*University of Guelph*
Guelph, Canada
hkarimi@uoguelph.ca

*Abstract*—Smart power grids are one of the most complex cyber-physical systems, delivering electricity from power generation stations to consumers. It is critically important to know exactly the current state of the system as well as its state variation tendency; consequently, state estimation and state forecasting are widely used in smart power grids. Given that state forecasting predicts the system state ahead of time, it can enhance state estimation because state estimation is highly sensitive to measurement corruption due to the bad data or communication failures. In this paper, a hybrid deep learning-based method is proposed for power system state forecasting. The proposed method leverages Convolutional Neural Network (CNN) for predicting voltage magnitudes and a Deep Recurrent Neural Network (RNN) for predicting phase angels. The proposed CNN-RNN model is evaluated on the IEEE 118-bus benchmark. The results demonstrate that the proposed CNN-RNN model achieves better results than the existing techniques in the literature by reducing the normalized Root Mean Squared Error (RMSE) of predicted voltages by 10%. The results also show a 65% and 35% decrease in the average and maximum absolute error of voltage magnitude forecasting.

## I. INTRODUCTION

In smart power grids, it is critically important to have accurate information about the current state of the system and the state variation tendency [1], [2]. State estimation is extensively employed in energy management systems for monitoring of the smart power grid state [3]–[5]. It is used to retrieve the unknown system state, including voltage magnitudes and phase angles using a set of measurements provided by the Supervisory Control and Data Acquisition (SCADA) system [6], [7]. However, state estimation is highly sensitive to measurement corruption due to the bad data or temporary malfunction of the communication system [8], [9]. Therefore, state forecasting is adopted to address the issue by predicting system states ahead of time, which in turn enhances the state estimation accuracy and security [8], [10], [11]. State forecasting aims to produce sufficiently accurate forecasts for voltage at all buses of the system. The forecasted quantities are mostly utilized for debugging incoming data obtained from the SCADA system [8].

A considerable number of studies have been carried out on power system state forecasting since the 1970s, which can be divided into two categories: traditional methods and learning-based methods. The first research in this field was presented by Debs in [12], who introduced the idea of the tracking state estimator. The Kalman filtering technique was adopted in the study above and an identity matrix was used for the state transition matrix. Nishiya in [13] improved the state forecasting by introducing a trend factor to the model. However, the state transition matrix was still a unity matrix. The model was further improved by estimating the diagonal elements of the transition matrix by the Holt's Exponential Smoothing (HES) method. Morvaj also used Kalman filter [14] for the power system dynamic state estimation, but the elements of the diagonal transition matrix were updated hourly based on load forecasting. A power flow-based dynamic model driven by load forecasts was presented in [15], which brings in a new term to the state transition model. A block-diagonal state transition matrix based on regression analysis was presented in [1] for power system state forecasting. In this study, updates occur in elements of the transition matrix, when new measurement data is received. Another approach was proposed by Hassanzadeh in [3] for short-term nodal voltage phasor forecasting using first-order vector autoregressive (VAR) modeling. However, the assumption of linear dynamics in all of these methods is practically unrealistic, since there is nonlinear relationship between system states.

Artificial Neural Networks (ANNs) have also been used in power system state forecasting. ANN was utilized by Filho in [16] in order to achieve state forecasting. However, the adopted architecture corresponds to a linear model. The ANN-based state forecasting model presented in [8], [17] is also the same proposed in [16]. The state transition matrix elements correspond to the ANN interconnection weights. The drawback of this model is that the model parameters increase with the length of input sequences. The study presented in [18] utilizes ANN for predicting bus loads in the prediction stage of the dynamic state estimation. Then, power flow equations were adopted to determine voltage phasors. A Deep Recurrent Neural Network (RNN) was proposed in [10] for power system state forecasting. The RNN model outperforms existing alternatives by considering nonlinear dependencies between system states. Besides, contrary to the study presented in [17], the number of model parameters for variable-length input sequences is fixed.

In this paper, a hybrid deep learning-based method is proposed for power system state forecasting in order to achieve high prediction accuracy. The proposed method consists of Convolutional Neural Network (CNN) for predicting voltage magnitudes and a deep RNN for predicting phase angels. The reason for proposing the CNN model for predicting voltage magnitudes is that CNN is specialized in processing data with a grid-like topology that fit our smart power grid data. The proposed method called CNN-RNN model is evaluated on the IEEE 118-bus benchmark, and the results have been compared with the RNN model in [10], which has the best performance among the existing alternatives in the literature. The results demonstrate that the CNN-RNN model improves prediction accuracy. The normalized Root Mean Squared Error (RMSE) of predicted voltages reduced by 10%. Besides, the average and maximum absolute error of voltage magnitude forecasts decreased by 65% and 35%, respectively.

This paper is organized as follows: Section II presents the power system state forecasting. The proposed CNN-RNN model is described in Section III. Section IV provides simulation results and discussions. Finally, the paper is concluded in Section V.

## II. POWER SYSTEM STATE FORECASTING

The purpose of power system state forecasting is to predict the next state of the system accurately $S_{t+1}$ using the previous system states $\{S_\tau\}_{\tau=t-r+1}^{t}$ as expressed in (1) [10]:

$$S_{t+1} = f_{state\ forecasting}(S_{t-r+1}, \dots, S_{t-1}, S_t) + \xi_t \quad (1)$$

where $S_t$ is the state of the system at time instance $t$. The number of lagged states are represented by the parameter $r \geq 1$. Function $f_{state\ forecasting}$ is an unknown nonlinear function that forecasts the next state $S_{t+1}$ using a lagged system states $\{S_\tau\}_{\tau=t-r+1}^{t}$. In addition, $\xi_t$ models inaccuracies.

The state of the system is represented by bus voltages (magnitudes and phase angles) at all buses [8]. The number of state variables is twice the number of buses ($n$), which includes $n$ voltage magnitudes and $n$ phase angles. State variables at time instance $t$ are shown in (2):

$$S_t = [|V_t^1|\ |V_t^2|\ \dots\ |V_t^n|\ \angle V_t^1\ \angle V_t^2\ \dots\ \angle V_t^n]_{1\times 2n}^T \quad (2)$$

where $|V_t^n|$ and $\angle V_t^n$ represent voltage magnitude and phase angle at time instance $t$ and bus $n$, respectively.

## III. PROPOSED HYBRID DEEP LEARNING-BASED MODEL

The architecture of the proposed CNN-RNN model is illustrated in Fig. 1. The model consists of a CNN model developed for predicting voltage magnitudes and a deep RNN model developed for predicting phase angles. The upper half uses a one-dimensional (1D) CNN model for the multivariate time-series forecasting, followed by dense, fully connected layers. The lower half uses a three-layer deep RNN followed by a dense, fully connected layer. Model inputs consisting of state variables from time instance $t$-$r$+1 to $t$, are shown in green circles. According to (1) and (2), The input can be expressed as matrix of state variables (voltages) as presented in (3):

$$Input = \overbrace{[S_{t-r+1}\ S_{t-r+2}\ \dots\ S_{t-1}\ S_t]}^{r\ Lagged\ System\ States}$$

$$= \begin{bmatrix} |V_{t-r+1}^1| & |V_{t-r+2}^1| & \dots & |V_{t-1}^1| & |V_t^1| \\ |V_{t-r+1}^2| & |V_{t-r+2}^2| & \dots & |V_{t-1}^2| & |V_t^2| \\ \vdots & \vdots & \ddots & \vdots & \vdots \\ |V_{t-r+1}^n| & |V_{t-r+2}^n| & \dots & |V_{t-1}^n| & |V_t^n| \\ \angle V_{t-r+1}^1 & \angle V_{t-r+2}^1 & \dots & \angle V_{t-1}^1 & \angle V_t^1 \\ \angle V_{t-r+1}^2 & \angle V_{t-r+2}^2 & \dots & \angle V_{t-1}^2 & \angle V_t^2 \\ \vdots & \vdots & \ddots & \vdots & \vdots \\ \angle V_{t-r+1}^n & \angle V_{t-r+2}^n & \dots & \angle V_{t-1}^n & \angle V_t^n \end{bmatrix}_{2n\times r} \quad (3)$$

Similarly, the output of the model, which is shown in blue circles, can be expressed as a vector of state variables in (4):

$$S_{t+1} = [|V_{t+1}^1|\ \dots\ |V_{t+1}^n|\ \angle V_{t+1}^1\ \dots\ \angle V_{t+1}^n]_{1\times 2n}^T \quad (4)$$

### A. Convolutional Neural Network (CNN) Model

CNNs have a great performance in processing data with grid-like topology. Therefore, they can be adopted for time series forecasting by considering time series data as a 1D grid, which takes samples at regular time intervals [19].

The input to the CNN model is the $2n \times r$ matrix of state variables presented in (3). In this paper, $n$ and $r$ account for 118 buses and 10 lagged states, respectively. As shown in Fig. 1, the first stage of the CNN model is a convolutional layer. In this layer, 118 filters with the kernel size of 2 are applied to the matrices of $2n \times 2$ dimension given in (5) to produce feature maps. After that, feature maps pass through a nonlinear Relu function. The equation describing the convolutional layer is presented in (6).

$$I_p = [S_{t-p}\ S_{t-p+1}]_{2n\times 2} \quad 1 \leq p \leq r-1 \quad (5)$$

$$h_p^k = a_{Relu}(w^k * I_p + b^k) \quad 1 \leq p \leq r-1,\ 1 \leq k \leq 118 \quad (6)$$

where $a_{Relu}$ is the activation function, which is selected as a Relu function in this paper. The matrix $w^k$ and the scaler $b^k$ are the weight matrix and bias related to the $k$th filter, respectively. Also, $h_p^k$ stands for the $p$th element of the $k$th feature map. Finally, $I_p$ represents the $p$th region of the input to which the filter is applied.

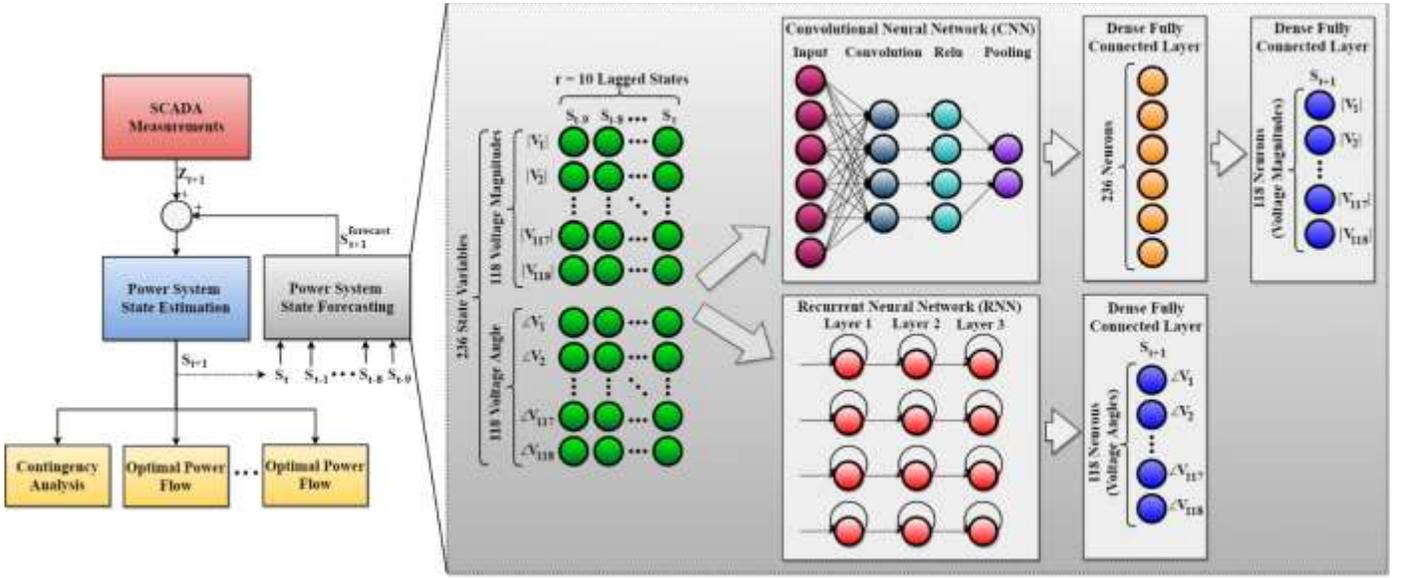

Fig. 1. The architecture of the proposed CNN-RNN model for power system state forecasting.

The convolutional layer is followed by a pooling layer. A pooling layer receives input and changes the elements of the input by a summary statistic of its nearby elements [19]. Max pooling operation is used in the pooling layer of this model. The equation expressing max-pooling operation is presented in (7):

$$P^k = max\ pooling(h_p^k) \quad 1 \le p \le r-1,\ 1 \le k \le 118 \quad (7)$$

where $P^k$ is the output of the pooling layer corresponding to the $k$th feature map. The pool size is selected to be 2; therefore, $P^k$ is a vector with 4 elements for each feature map.

The pooling layer is followed by a flatten layer, which converts the output of the pooling layer with a $4 \times 118$ dimension to a vector with 472 elements, as expressed in (8). The next stage is a dense, fully connected layer with 236 neurons. The output of the dense layer is passed through a Relu activation function, as presented in (9).

$$o_{flat} = Flatten(P) \quad (8)$$

$$o_{dense1} = o_{Relu}(W_{dense1}\ o_{flat}) \quad (9)$$

where $o_{flat}$ and $o_{dense1}$ are the output vectors of the flatten and dense layers, respectively. Also, $W_{dense1}$ is the weight matrix corresponding to the dense layer.

The final layer of the CNN architecture is another dense, fully connected layer with 118 neurons. The output of the second dense layer is passed through a linear activation function to produce the outputs of the CNN model, which are voltage magnitudes of all busses as presented in (10):

$$[|V_{t+1}^1|\ |V_{t+1}^2|\ \cdots\ |V_{t+1}^n|]_{1\times n}^T = o_{Linear}(W_{dense2}\ o_{dense1} + b_{dense2}) \quad (10)$$

where $W_{dense2}$ and $b_{dense2}$ are the weight matrix and bias vector of the second dense, fully connected layer, respectively.

### B. Deep Recurrent Neural Network (RNN) Model

RNNs, as a member of the neural network's family, are specialized to process sequential data having variable-length inputs or outputs [19], [20]. In principle, recurrent networks can use their feedback connections to store representations of recent input events in the form of activations (short-term memory). This is potentially significant for many applications, including speech processing, non-Markovian control, music composition, and natural language processing (NLP) [21].

A dynamical system influenced by an external signal can be described by (11) [19]:

$$S_t = f_\theta(S_{t-1}, x_t) \quad (11)$$

where $S_t$ and $x_t$ account for the system state and the external input at time instance $t$, respectively. The function $f_\theta$, which includes parameters $\theta$, is responsible for mapping $S_{t-1}$ to $S_t$.

The equation expressed in (11) can be visually represented in Fig. 2. The left half and right half illustrate the circuit form and its unfolded flow graph, respectively. In the right half, each node indicates a time instance, while a delay of 1-time instance on the recurrent connection is demonstrated by a black square in the left half. A simple RNN architecture is shown in Fig. 3 [19]. The equations describing this architecture are expressed in (12) and (13) [22]:

$$S_t = f_h(x_t, S_{t-1}) = \emptyset_h(U\ x_t + W\ S_{t-1}) \quad (12)$$

$$o_t = f_o(x_t, S_t) = \emptyset_o(V\ S_t) \quad (13)$$

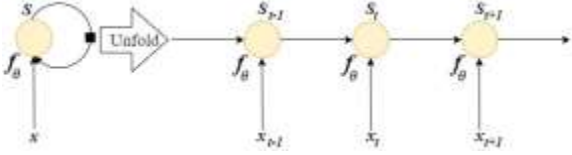

Fig. 2. Correlation between the current state and the whole past sequence.

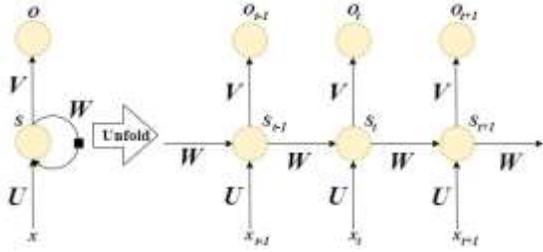

Fig. 3. A simple RNN.

where $U$, $W$, and $V$ are the weight matrices for input-to-hidden, hidden-to-hidden, and hidden-to-output connections, respectively. The input and output of the dynamical system are expressed by $x_t$ and $o_t$ respectively. The functions $f_o$ and $f_h$ stand for the output function and transition function. The functions $f_o$ and $f_h$ are defined by a set of parameters $\theta_o$ and $\theta_h$ respectively. The functions $\emptyset_o$ and $\emptyset_h$ account for element-wise nonlinear functions.

Deep RNNs are constructed based on RNN architecture with multiple processing layers [10]. The deep RNN used in this paper is a stacked RNN, which includes multiple levels of transition functions, one on top of another [22]. The architecture of the deep RNN and its corresponding equations are presented in Fig. 4 and (14), respectively.

$$S_t^l = f\big(R_v^{l-1} S_t^{l-1} + R_h^l S_{t-1}^l + r^{l-1}\big) \tag{14}$$

where $l$ stands for the layer index, and $f$ is a nonlinear activation function (Relu in this paper). Also, $\{R_v^l, R_h^l, r^l\}$ are the weight matrices and vectors. $S_t^{l-1}$ is the hidden state at time $t$ and layer $l$-$1$.

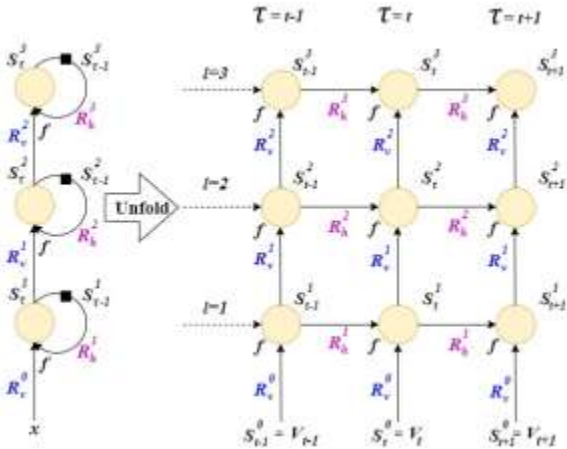

Fig. 4. The architecture of the deep RNN in the proposed method.

In this paper, the deep RNN has three layers ($l = 3$) and ten lagged states ($r = 10$). It is followed by a dense, fully connected layer with 118 neurons, as shown in Fig. 1. The equation demonstrating phase angle forecasting is presented in (15).

$$[\angle V_{t+1}^1 \quad \angle V_{t+1}^2 \quad \cdots \quad \angle V_{t+1}^n]_{1 \times n}^T = o_{Linear}\big(W_{dense3}\, S_t^l + b_{dense3}\big) \tag{15}$$

where $\angle V_{t+1}^n$ stands for the phase angle forecast at time $t+1$ and bus $n$. In addition, $W_{dense3}$ and $b_{dense3}$ represent the weight matrix and bias vector of the dense fully connected layer.

## IV. SIMULATION RESULTS AND DISCUSSION

The proposed CNN-RNN model developed for power system state forecasting is evaluated on the IEEE 118-bus benchmark. The dataset, which consists of 236 state variables (bus voltages, including 118 voltage magnitude and 118 phase angles) at 18,528 time instances, is adopted from [10]. The training set and test set include 80% and 20% of the dataset with 14,822 and 3,706 samples, respectively. The dataset is developed based on the real load data of the 2012 Global Energy Forecasting Competition (GEFC) [23]. The simulation results of the proposed CNN-RNN model are compared with the deep RNN model presented in [10], which outperforms existing techniques in the literature. The efficient Adam version of stochastic gradient descent is used to fit the model and learn weights. The model is implemented using Python programming language in a system with Intel(R) Core(TM) i5-7400 CPU with 16 GB RAM specifications.

The CNN-RNN and RNN models are evaluated in terms of normalized RMSE. The RMSE of the proposed model is computed by determining the average RMSE over 20 independent runs. The average RMSE of the CNN-RNN and RNN models are $2.331 \times 10^{-3}$ and $2.588 \times 10^{-3}$ respectively. Therefore, the proposed model reduces the RMSE by 10% and improves the prediction accuracy.

The reason for the superiority of the proposed model is using the CNN architecture for predicting voltage magnitudes. CNNs specialize in processing data with grid-like topology and mining spatial features that fit our smart power grid data. The average and maximum Absolute Errors (AE) of voltage predictions over the whole test set are presented in TABLE I. A 65% and 35% decrease in the average and maximum AEs of voltage magnitudes are observed, respectively. It is clear that the proposed CNN-RNN model significantly reduces the voltage magnitude prediction error.

It is also observed that the voltage angle prediction error is almost the same in both cases. The reason for the same results is the use of three layers deep RNN in both methods for forecasting phase angles. The results are confirmed by plotting AE over the whole test set. The AEs of voltage magnitude and voltage angle forecasts are demonstrated in Fig. 5, and Fig. 6, respectively.

TABLE I. ABSOLUTE ERROR OF VOLTAGE FORECASTING

| Method | Absolute Error of Voltage Magnitude (p.u.) | | Absolute Error of Voltage Angle (degree) | |
|---|---|---|---|---|
| | *Average* | *Max* | *Average* | *Max* |
| CNN-RNN | $9.87 \times 10^{-4}$ | $1.49 \times 10^{-2}$ | 1.20 | 10.63 |
| RNN | $2.80 \times 10^{-3}$ | $2.31 \times 10^{-2}$ | 1.35 | 10.62 |

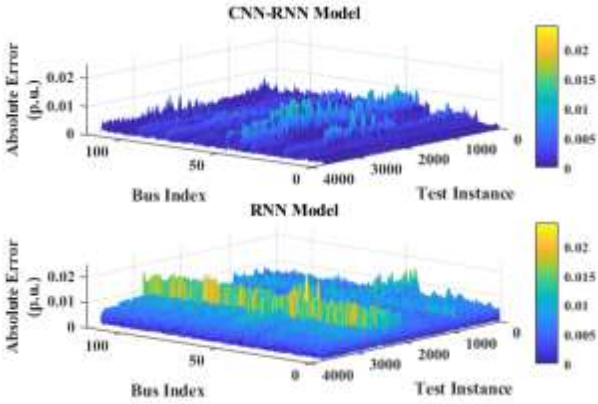

Fig. 5. The absolute error of voltage magnitude forecasting for the CNN-RNN and RNN models.

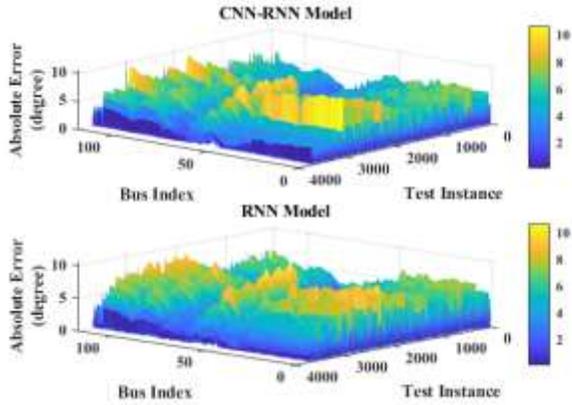

Fig. 6. The absolute error of voltage angle forecasting for the CNN-RNN and RNN models.

Larger errors are evident in forecasted voltage magnitudes by the RNN model compared to the proposed CNN-RNN model in Fig. 5.

The forecasted voltages for all buses at test instance 3222 is shown in Fig. 7. The proposed CNN-RNN model predicts voltage magnitudes with considerably better accuracy than the RNN model. However, due to the use of the same three-layer deep RNN in both methods for phase angle forecasting, the accuracy of both models in predicting phase angle is the same.

Also, Fig. 8 illustrates voltage magnitude and angle forecasts for bus 16 from test instance 500 to 1000 (It is noteworthy that the test instance numbering starts from the 11th test example since $r=10$). The figures clearly show that the proposed CNN-RNN model considerably improves the accuracy of the voltage magnitude predictions, while the phase angle forecasting accuracy is almost the same.

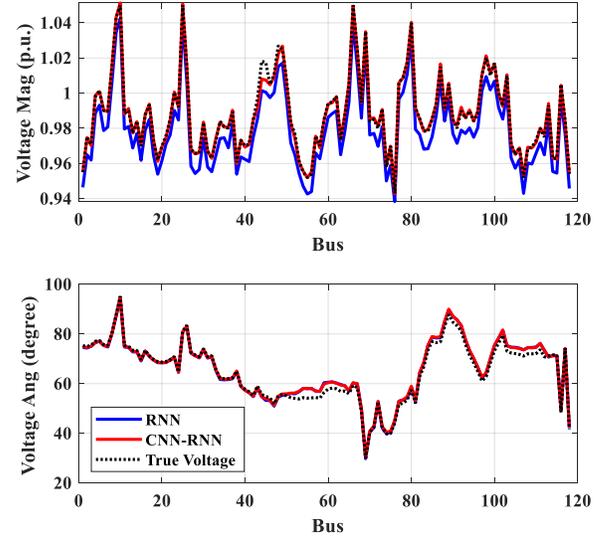

Fig. 7. Forecasted voltage magnitudes and angles for all buses at test instance 3222.

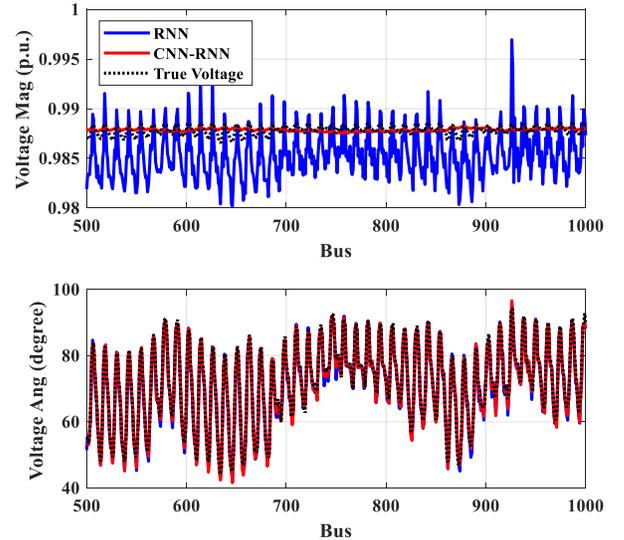

Fig. 8. Forecasted voltage magnitudes and angles of bus 16 from test instance 500 to 1000.

## V. CONCLUSION

A hybrid deep learning-based method was proposed for power system state forecasting. The proposed method consists of a CNN model for predicting voltage magnitudes and a deep RNN model for predicting voltage angles. The CNN-RNN model produced more accurate predictions by reducing the

RMSE of forecasted voltages by 10% compared to the best existing technique in the literature. In addition, using CNN for voltage magnitude forecasting considerably decreased the average and maximum absolute errors of voltage magnitude forecasts by 65% and 35%, respectively. Therefore, CNNs can be a better option than other alternatives for voltage magnitude forecasting. The higher accuracy achieved by CNNs is due to the fact that they specialize in processing data with grid-like topology and mining spatial features that fit our smart power grid data. The contribution of our future work will be adopting series and parallel CNN-RNN architectures to improve power system state forecasting accuracy further.